\documentclass[conference]{IEEEtran}
\usepackage{comment}
\usepackage{wrapfig}
\usepackage{setspace}
\usepackage{hyperref}
\usepackage{url}

\begin{document}
\title{Run-time extensibility and librarization of simulation software}
\author{
\IEEEauthorblockN{Jed Brown\footnote{Corresponding author}}
\IEEEauthorblockA{Argonne National Laboratory\\
University of Colorado Boulder \\
jedbrown@mcs.anl.gov}
\and
\IEEEauthorblockN{Matthew G.~Knepley}
\IEEEauthorblockA{University of Chicago\\
knepley@ci.uchicago.edu}
\and
\IEEEauthorblockN{Barry F.~Smith}
\IEEEauthorblockA{Argonne National Laboratory\\
bsmith@mcs.anl.gov}
}
\maketitle

\begin{abstract}
  Build-time configuration and environment assumptions are hampering progress and usability in scientific software.
  That which would be utterly unacceptable in non-scientific software somehow passes for the norm in scientific packages.
  The community needs reusable software packages that are easy use and flexible enough to accommodate next-generation simulation and analysis demands.
\end{abstract}

I'd like you to use my new web browser, Firetran!
It renders HTML 10\% faster than Firefox, but only if there is no JavaScript.
You can recompile if you want JavaScript, but we don't show that in our performance tests.
The character encoding is compiled in, for efficiency.
It has a great plugin community---developers add code directly to the web browser core, guarded by a \texttt{\#ifdef}.
Some developers change things that I don't like, so they distribute their own version of Firetran.
Naturally, users of those packages submit bug reports to me and I ignore them because I can't reproduce with my version.
Proxy configuration is compiled in so you don't have to worry about run-time configuration dialogs, just edit a makefile and recompile.
To keep you secure, the https version of Firetran cannot use http and vice-versa.
Although Firetran is open source, our development is done in private, so if you submit a bug report or a patch, you'll likely hear ``We fixed that in the private repository last year; we'll release when the paper comes out.  If you have to view that website, fill out the attached form and fax us a signed copy.''
Firetran has a parental filter feature: you can list a maximum of 16 websites in a source file, in which case Firetran will refuse to visit any site not on the list.
Firetran can only be compiled with last year's version of the ACME Fortran77 compiler.
The build system consists of csh, perl, m4, and BSD make.
There is no URL entry box in Firetran; to visit a page, you edit a configuration file and run the program.
A grad student wrote a Tcl script with a text entry box to automate editing the configuration file and re-running the Firetran executable.
The script is hard to understand, but many in the community believe the way forward is to enhance the script to detect whether the website needs https or http, JavaScript or not, etc., and recompile Firetran on the spot.

Needless to say, Firetran struggles with market share.
And yet choices in Firetran represent the status quo in many scientific software packages, often vehemently defended.
If it is laughably unacceptable in non-scientific software, why is it tolerated in scientific software?
Are scientists suffering from Stockholm Syndrome?
Is scientific software so fundamentally different?
How could scientific software benefit from adopting the techniques we take for granted in non-scientific software?
Let's look at some of the directions that scientific simulation software is headed.

\section{Trends in simulation-based science and engineering}
Modern computational science and engineering is increasingly defined by multiphysics, multiscale simulation~\cite{KeyesMcInnesWoodwardEtAl13} while raising the level of abstraction to risk-aware design and decision problems.
This evolution unavoidably involves deeper software stacks and the cooperation of distributed teams from multiple disciplines.
Meanwhile, each application area continues to innovate and can be characterized as much by the forms of extensibility (e.g., boundary conditions, geometry, subgrid closures, analysis techniques, data sources, and inherent uncertainty/bias) as by the underlying equations.
It is no longer the case that the original author can foresee all use cases for their software.
We argue that many common approaches to configuration and extensibility create \emph{artificial bottlenecks} that impede science goals, and that the only sustainable approach is to defer these to run-time.
Doing this effectively will push applications to minimize the assumptions made about their environment, resulting in more library-like applications better-suited to coupling with other models and to performing advanced analysis.

\subsection{Compile-time configuration}
The status quo for many applications, especially those written in Fortran~\footnote{A language that after more than 50 years, has finally started to provide mechanisms for encapsulation in its latest standards (using ISO C bindings) and natively in TS 29113 (scheduled for inclusion in Fortran2015).}, is to perform configuration in the build system.
This is motivated by a variety of concerns about efficiency (often ill-founded or fixable by adjusting interface granularity), limitations of software tools (e.g., algorithmic differentiation), poor language support, perceived implementation complexity, and short-term value assessment.
Once a package chooses compile-time configuration, the build system becomes a public application programming interface (API), used by scripts that perform higher-level analysis.
Ad-hoc public APIs inhibit software evolution by imposing an unintentionally-high cost to change and dilution of effort to meet short-term deliverables.

In applications relying on build-time code generation, pragma-based specialization/optimization, or those written in C++ with heavy use of templates, the possible combinations must be enumerated at compile-time.
Although templates are not exclusive (you can compile several variants in the same application), it is common to see a combinatorial explosion of variants as well as directly exposing the templates in public interfaces.
Since all combinations cannot be compiled into one application, the effect is that any analysis or testing that explores a large or unpredictable part of the space of combinations must include recompilation.
Attempting to push the size limits leads to error-prone workarounds like \texttt{-mcmodel=large} (a compiler option that affects linking/compatibility), using processes spanning more than one NUMA node (degrading memory locality), and inability to run on low-memory architectures that might otherwise be well-suited to the application.

Compute nodes often do not have access to compilers, making all build-system and compile-time decisions inaccessible to online analysis.
It may be impossible for the same application to run in both configurations on different nodes or on different MPI communicators.
This limits analysis capability, requires frequent recompilation, and increases the likelihood of user error resulting from accidentally using the wrong compiled version.
The length of batch queues exacerbates the issue, sometimes requiring days between compiling an application and actually running it.
Every compatibility that must be maintained by hand is another opportunity for mistakes, some of which the user may not realize prior to publication.

Some applications create sophisticated scripts for maintaining consistency through the compilation and batch submission process.
These scripts must be ported to each architecture, and increase the complexity of debugging the application and of reproducing problems encountered on certain architectures.

Integration tests often need to be submitted to batch systems.
If different integration tests need dependencies to be compiled differently, those different versions need to be built in advance and kept straight through the test submission and run.
When many configurations are needed, the multiple required compilations have a tendency to take a long time and/or burn through disk quota.

\subsection{Advanced analysis}
As models mature in each application area, emphasis shifts from qualitative and subjective interpretation of model output to quantitative analysis of accuracy, reliability, and the influence of parameters on quantities of interest.
Correspondingly, today's models are increasingly used not just as forward models but as the target of advanced analysis techniques such as stochastic optimization, risk-aware decisions, and stability analysis.
The forward model must then expose an interface for each form of modification that the analysis levels can explore.
An interface requiring build-time modification shifts an unacceptable level of complexity to the analysis software and is algorithmically constraining---limiting parallelism, introducing artificial bottlenecks, and preventing some algorithms.

In lieu of tractable deterministic techniques for calibration of empirical phenomenological models, an enormous amount of expert time must be spent tuning parameters.
In fields such as climate, earthquakes, and molecular dynamics, this calibration is notoriously sensitive to numerical methods, temporal and/or spatial resolution, and other models used in simulation.
And yet when faced with this extreme uncertainty and volatility, these parameters are often hard-coded in the source, thwarting reasonable attempts to automate the calibration or comparison of models.

\subsection{Model coupling}
A large fraction of successful scientific software has been the result of a visionary scientist operating in a single domain.
Many important model configurations and analysis types were predicted by that visionary, and the community has been largely content to explore within that fuzzy scope.
Each package has been king of its own environment, thus could make choices without concern for interoperability or impact on other packages.
But the gaping holes in our scientific understanding and engineering capability lie increasingly in the gaps not covered by any one of these mature packages.
Rarely do multiple models operate on identical spatial and temporal scales with similar model and parameter uncertainties, thus coupling often requires grappling with multiscale phenomena and high-variance statistics, each algorithmic challenges in their own right.
When components make excessive assumptions about their environment, attempts to couple are either written off or algorithmic quality falls by the wayside, leading to nominally-coupled simulations that are unreliable at best and effectively non-convergent in most cases.
The most powerful and pragmatic software approach we know of is to formulate models as libraries with a clean hierarchy of interfaces, allowing an external client to compose the key capabilities into a coupled model without the higher-level parts that would algorithmically constrain a coupled model.
This approach has repeatedly demonstrated its effectiveness outside of scientific computing in areas traditionally dominated by stand-alone applications, such as compilers (LLVM), web browsers (KHTML/WebKit), and SQL databases (SQLite).
Although process isolation can be useful for reasons of security (e.g., qmail, postfix), reliability (tabs in a web browser), and distribution (e.g., remote databases), it is easier to add isolation upon library interfaces than to add composition/embedding atop process separation, especially in HPC environments for which oversubscription is usually catastrophic.

\subsection{Provenance and usability}
Reproducibility and provenance are perpetual challenges of computational science that become more acute as the software stack becomes deeper and more models of greater complexity are coupled.
How can we capture the state of all configuration knobs so that a computational experiment can be reproduced?
Compare the complexity of a single configuration file to be read at run-time with that of a heterogeneous configuration consisting of multiple build systems, files passed from earlier stages of computation, and run-time configuration.
Provenance is simplified by using each package without modification, compiled in a standard way, and controlled entirely via run-time options.
This implies that any libraries used (transitively) by the application must be responsible libraries that adhere to the principles discussed here and in \cite{gropp98}.
For both maintenance and provenance reasons, custom components needed for a given computational experiment are better placed in version-controlled plugins rather than by modifying upstream sources.
If a coherent top-level specification is to be supported in a system with build-time or source-level choices, those configuration options must be plumbed through all the intermediate levels, often resulting in another layer of ``workflow'' scripts and bloated, brittle high-level interfaces.


\subsection{``Big'' data}
Workflows that involve multiple executables usually pass information through the file system.
It takes about one hour to read or write the contents of volatile memory to global storage on today's top machines, assuming peak I/O bandwidth is reached.
The largest allocations are on the order of tens of millions of core hours (e.g., INCITE), meaning that the entire annual compute budget can be burned in a few reads and writes.
Global storage as an \emph{algorithmic} mechanism is dead: where out-of-core algorithms have been used in the past, today's scientists can simply run on more cores, up to the entire machine; but if the entire machine does not have enough storage, the allocation simply does not have the budget to run an out-of-core algorithm.

If a different application or different version of an application must be used for the next stage in the simulation/analysis pipeline, data must be dumped to the file system.
In-situ analysis provides an excellent opportunity to increase efficiency by reducing dependence on the file system, but is only viable if the more varied analysis workflow can be performed in the same application.
Interfaces for exchanging data in-memory between different software components could be the same as those used to describe data sets for parallel IO.

Some of today's simulations support a large and diverse community that analyzes the output.
Transitioning to in-situ analysis will require dynamic and extensive analysis interfaces to support varied analysis demands.
Unlike most parts of mature simulation software, the analysis code often changes with each question a scientist asks, thus is highly volatile and does not benefit from the same amount of testing.

\subsection{Nested dependencies}\label{sec:nested}
Some library dependencies are indirect (transitive), via some intermediate interface that the application actually intends to depend on.
One of the most important software engineering principles is that of encapsulation, allowing clients to depend only on interfaces that it uses directly, rather than implementation concerns.
There is no encapsulation if a transitive dependency must be reconfigured for each use case, and combining uses into one application may cause conflicts.
The build system for any ``library'' that requires use-specific configuration effectively becomes a public API that top-level components must interact with even when the library is only used indirectly.

A single library can be used by multiple components in the same executable.
This may be rare when a library is first being developed, but is common among popular and versatile libraries.
If a library has mutually-incompatible configurations, the entire executable can only use one version unless the library developer has taken great care (often impractical, especially when linking statically---an unfortunate necessity on many HPC architectures).
Even in the best case, needing to use multiple versions complicates the installation and debugging process, invariably leading to degraded user experience and increased support workload for library maintainers.

\subsection{User modifications}
Fragmentation of software projects is notoriously expensive and should be avoided when possible.
Maintaining local modifications with no plan for upstreaming is a recipe for divergent design---technical debt that must be paid off to combine the features developed in each fork.
Fragmentation is especially toxic for libraries that may be used by multiple higher-level packages that are combined in the overall experiment (see \autoref{sec:nested})

\subsection{Packaging and distribution}
Software developers often underestimate the challenge of installing their own packages.
From the perspective of user experience, it hardly matters if an installation failure was caused by a user's broken environment (a circumstance all-too familiar to maintainers of popular packages).
Upgrading an operating system can break existing installs of a package if the underlying system libraries change.
The most reliable way to distribute packages that will always be in-sync with the operating system is to have them be packaged by many common operating systems (Debian APT, RedHat RPM, MacPorts, etc.).
Configure-time options are the bane of package distribution due to the need to name each variant and to resolve conflicts between the variants.
Packagers for binary distributions (most convenient for users) are justifiably paranoid about the binary interface, so will be reluctant to package software with fragmented configuration options.

\section{Implementation and recommendations}
To manage these workflow challenges, application developers will need to think more like library developers~\cite{gropp98}, controlling namespaces, avoiding global state, relinquishing top-level control, controlling the scope of parallelism, localizing memory allocation, localizing complexity so that it does not ``bubble up'' to the top level, and paying attention to the completeness, generality, stability, and extensibility of all public interfaces.
Our suggestions are shaped by experience developing and supporting (\href{http://mcs.anl.gov/petsc}{PETSc})~\cite{petsc-user-ref,petsc-developers}, as well as other packages from low-level libraries to end-user applications.
Similar ideas for extensibility and run-time configuration have been implemented in applications such as MOOSE~\cite{moose-web-page} and PyLith~\cite{pylith-web-page}.

\subsection{Resource allocation}
To localize configuration, allocation of resources such as memory should be done locally, with reference counting when appropriate.
Contrary to urban legend, static memory allocation offers no tangible performance advantage (so long as dynamic allocations are amortized) and unavoidably ties the workflow into the build system, while committing the sin of needless global variables.
Different \texttt{malloc} implementations have varying performance, especially in multi-threaded scenarios.
If necessary, fast implementations like TCMalloc~\cite{ghemawat2009tcmalloc} can be recommended, but it is better to contain this complexity in order to perform well with any \texttt{malloc}.
This can involve having memory pools or work arrays associated with algorithm objects, so that \texttt{malloc} is not called in inner loops.

\subsection{Plugins}
Source-level dependencies on an implementation (e.g., direct instantiation of a derived class or a template parameter) rather than a generic interface cause choices from deep in the stack to ``bubble up'' via brittle interfaces that plumb the user's configuration to the appropriate component.
Plugins provide a strong way to identify interfaces that can be extended by users and distributed separately from the core package.
For example, every class in PETSc has a plugin architecture, from base linear algebra components to preconditioners, nonlinear solvers, and adaptive controllers for time integration.
Any of these components can be provided by a plugin and will be indistinguishable from a native component of PETSc.
Plugins consist of a registration function that is called via \texttt{dlopen()}, a creation function that is called when the plugin is activated (e.g., instantiation of an object implemented in the plugin), and any supporting functions that will be exposed via methods of the object.
Historically, Fortran's type system and inability to store function pointers have conspired against plugin implementations, but the new standard provides the necessary tools.

Plugins also provide a mechanism to invert dependencies without creating dependency loops.
For example, suppose \texttt{libB} depends on \texttt{libA}, but we would like to provide an optional implementation of an interface in \texttt{libA} that depends on \texttt{libB}.
We can't put it in \texttt{libA} because this would make a cyclic dependency, but it is unrelated to \texttt{libB}'s public interface so doesn't belong there either.
We \emph{can} create \texttt{libA-plugin} that depends on both libA and libB, registering itself as a plugin of \texttt{libA} and calling into \texttt{libB} in its implementation.
Note that plugins can also be used for optional interfaces to third-party libraries.
It is best to have plugin search paths from which plugins are loaded by \texttt{dlopen}, so that they can be distributed independently from the base system and no relinking is required.
Shared libraries should be versioned (e.g., \texttt{-soname} on most POSIX systems, \texttt{-current\_version} and \texttt{-compatibility\_version} on OSX) to make this distribution more reliable and to assist the layers built on top.
See \cite{drepper2002dsohowto} for more on shared library versioning and controlling symbol visibility.

While distribution via shared libraries is convenient for users and packagers, some important HPC execution environments do not support shared libraries.
If such anti-productive environments must be used, the plugin structure can be preserved, but the build system ultimately needs to be able to link everything statically.
For an application, this typically means that plugin source trees are placed in a location that the build system picks up, then code to call the registration function is generated, and everything is linked together.
For a library, plugins either need to be compiled into a single static archive or the user needs to explicitly link the plugins (in the correct order).
The linking interface is a public interface, so changing it should not be taken lightly.
The library can either distribute a tool that determines which plugins are available and generates a suitable link line or it can create a static archive containing all plugins.
Unfortunately, the \texttt{pkg-config} tool is lacking in management of multiple configurations and optional dependencies, so many libraries will need their own executable.
Wrapper compilers are exclusive (only one library can use a wrapper compiler) and thus should be avoided.

\subsection{Inversion of control, recursive configuration, and the options database}
The primary purpose of software libraries is to contain complexity.
Public interfaces should be as simple as possible (but no simpler), meaning that transitive complexity \emph{must not} be a mandatory part of the public interface.
Furthermore, extensible components are not known at compile-time (indeed, they may not have been written yet), thus would be rendered useless if implementation complexity leaked into the public interface.
It should be possible to instantiate the same plugin (unknown to client code) with different configurations at different locations in the object graph, each with its own configuration.
Since the client does not know how to configure the object, some inversion of control~\cite{fowler2004injection} is necessary.
PETSc's approach is similar to ``service locator'' in \cite{fowler2004injection}, but several variations should be considered by new projects.
In PETSc, multiple instances of objects are distinguished by a \emph{prefix} in the options database, allowing conflict-free run-time configuration.
For example, a multiphysics solver may use a block decomposition and geometric-algebraic multigrid with choices and diagnostics for each block and at each level of one or more multigrid solves, each instance of which we distinguish by prefix.
The basic principle is to choose good defaults and defer precise configuration to the run-time interface.
Some packages take dynamic extensibility further by embedding a Turing-complete programming language such as Lua, JavaScript, or Scheme.

Meanwhile, PETSc also acknowledges that some users take \emph{active} control over method configuration, adapting it in response to physical regime or other factors.
This active control is more naturally implemented and debugged with an object-based run-time interface, thus any run-time configuration exposed via the options database is also exposed via the object-oriented interface.
The most challenging compromise in this scenario occurs when an algorithm adaptively configures recursive levels, but the client wants to actively configure portions.
Solutions include fine-grained interfaces for ``forcing'' (in the lazy functional programming sense) certain parts of the setup and callbacks to configure portions when reached.
Neither are completely satisfactory.

\subsection{Object-oriented design}
We turn now to some contentious issues in object-oriented design, for which we are less than enamored with oft-repeated recommendations.

\subsubsection{Partial implementation}
Some people believe that all errors should be compile-time errors, thus any incompatibility must be visible to the compiler.
Unfortunately, this leads to extremely complicated and fragile type hierarchies.
For example, a \texttt{Matrix} is a linear transformation on finite-dimensional vector spaces.
Should a \texttt{Matrix} have computable entries?
Should the diagonal be extractable?
Can the transpose be applied?
Are ``Neumann'' sub-problems available (matrices with certain properties whose sum equals the original matrix)?
While in principle, the entries of matrices can be computed, the space and time complexity may be so unaffordable as to render that representation useless.
Meanwhile, other operations that are unaffordable for explicitly-stored matrices may be fast for matrices with special structure.
Different preconditioners (which may reside in plugins) may require different functional from the \texttt{Matrix}.
Any type system that can guarantee full implementation of a given \texttt{Matrix} interface will end up conflating the desired generic interface with implementation-specific semantics, especially when the \emph{Matrix} type is also extensible, leading to undesirable dependencies and leakage of transitive complexity.
Moreover, the ``not implemented'' run-time error is likely to be more understandable than a type mismatch error.

\subsubsection{Run-time implementation changing}
PETSc has found it useful for major objects to be able to change implementations at run-time (e.g., from multigrid to a direct solve).
One object can have many dependencies/references and be referenced by many other objects.
If the implementation can only be changed at object creation, the user ends up holding factory objects (or the equivalent) for the sole purpose of recreating ``similar'' objects.
Someone has to be responsible for keeping track of these factory objects and for rewiring the dependencies when replacing an existing object.
This turns out to be messy and error-prone, so PETSc has chosen to absorb the ``factory'' functionality into the object itself, allowing reconfiguration of any sort at any time.
This also removes the need for special interfaces to pass a factory object around to all components that should have a say in how the new object will be configured.

\subsubsection{Controlling the binary interface}
Time spent recompiling code is nothing but wasted productivity.
Implementation concerns such as private variables and new (virtual) methods should never require recompilation of client code.
PETSc uses a delegator (aka. ``pointer to implementation''~\cite{sutter2000exceptional} or ``bridge''~\cite{gamma1994design}) pattern to keep such implementation concerns out of the binary interface, thus minimizing recompilation and enabling binary distribution of shared library~\cite{drepper2002dsohowto} upgrades.
This is idiomatic in C where ``objects'' are usually implemented via opaque pointers, but often under-utilized in C++ because it entails a bit more boilerplate than the native object model that reveals the private contents of classes.
Delegator incurs an additional static function call, but tests with classic virtual methods and delegator indicate that the main function call overhead (several cycles) comes from the indirect call (virtual function) rather than the static call to the delegator, thus the incremental cost of using the delegator pattern is usually less than 2 cycles.
An ancillary benefit of the delegator pattern is that there is a unique place to set a debugging breakpoint for each function (rather than having to choose the correct virtual function) and a common place for input validation.

It is increasingly popular to expose libraries through more dynamic environments such as Python or Julia.
Since different languages have different type systems, it is easier and more reliable to develop language bindings with a simple type system and stable binary interface.
Naturally, static methods and opaque pointers are simpler than struct definitions and template-based systems.

\subsection{Just-in-time compilation}
In the case of fine-grained composition such as occurs in material models and Riemann solvers, as well as fusion of memory-intensive operations, the number of compositions grows combinatorially, but in any specific run, only a small number is important.
Precompiling and dispatching (via C++ templates or other inlining techniques) every combination leads to large compile times, bloated executables, confusing debugging, and compromises about which combinations will be made available.
While a dynamic interface is far more maintainable, the performance overhead is unacceptable for certain applications.
When the interface granularity cannot be increased to amortize the overhead of dynamicism, just-in-time (JIT) compilation is an attractive approach to preserve strong encapsulation and debuggability.
We anticipate technologies such as LLVM and OpenCL becoming ubiquitous, allowing judicious use of JIT for dynamic kernel fusion and plugin-style packaging of fine-grained components without sacrificing performance.
This may involve tighter integration with languages like Julia and the Numba package for Python, or language extensions to support JIT within traditionally compiled languages.

\subsection{Upstreaming, distribution, and community building}
In order to provide attractive alternatives to forking, maintainers must be diligent to create a welcoming environment for upstream contributions.
The maintainers should nurture a community that can review contributions, advise about new development approaches, and test new features, with recognition for all forms of contribution.
In a transparent community, it is immediately clear to paper reviewers who did the work to implement a new feature; thus any attempt to ``scoop'' a result based on new capability is easily spotted.
It is our opinion that scooping is a purely social problem and that the secrecy inherent in any technical solution bear a cost so expensive that it can rarely be justified.
Several major tech companies have famously underestimated this cost when forking open source packages such as the Linux kernel for internal use, later repaying the technical debt to reintegrate with upstream.
In science, it is exceedingly difficult to obtain funding to pay off the technical debt incurred by forking, leading to a wasteland of abandoned forks.
This is contrary to the interests of stakeholders ranging from the program managers and taxpayers to other scientists in the field.

In addition to community building~\cite{turk2013scale}, developers should provide versatile extension points so that contributions can be made without compromising existing functionality and without degrading package maintainability.
This should be thought of as a technical prerequisite for maintainable extension rather than private forking.
Such extensions must be accompanied by tests lest they break as interfaces evolve.
It is far easier to write tests for dynamic configuration sets than to add new build-time configurations.
Additionally, compilers and static analysis tools can check combinations that are not actively used, but conditional compilation is unchecked, invariably leading to more frequent breakage by other developers (in the test suite if covered, otherwise the breakage will be found by users and other developers).

\section{Conclusion}
Configuration and environment design decisions made by today's scientific libraries and applications are often disproportionately harmful to usability, productivity, and capability.
In such cases, the most effective way to increase scientific or engineering value is to design and refactor software using best practices for extensible library development.

\medskip

\subsection*{Acknowledgments}
JB and BFS were supported by the U.S. Department of Energy, Office of Science, Advanced Scientific Computing Research under Contract DE-AC02-06CH11357.
MGK acknowledges partial support from DOE Contract DE-AC02-06CH11357 and NSF Grant OCI-1147680.

\bibliographystyle{unsrt}
\bibliography{petscapp,petsc,jedbib}

\bigskip

\begin{quotation}
The submitted manuscript has been created by UChicago Argonne, LLC,
Operator of Argonne National Laboratory (``Argonne'').  Argonne, a
U.S. Department of Energy Office of Science laboratory, is operated
under Contract No. DE-AC02-06CH11357.  The U.S. Government retains for
itself, and others acting on its behalf, a paid-up nonexclusive,
irrevocable worldwide license in said article to reproduce, prepare
derivative works, distribute copies to the public, and perform
publicly and display publicly, by or on behalf of the Government.
\end{quotation}

\end{document}